\documentclass[a4paper,10pt,openany]{article}
\usepackage{pst-3d}
\usepackage{pst-3dplot}
\usepackage{etex}
\usepackage[utf8]{inputenc}
\usepackage[english]{babel}
\usepackage{amsmath,amssymb,amsthm,mathrsfs,amsfonts,dsfont}
\usepackage{graphicx}
\usepackage[footnotesize, skip=0pt]{caption}
\usepackage{epigraph}
\usepackage{indentfirst}
\usepackage{booktabs}
\usepackage{fancyhdr}
\usepackage{vmargin}
\usepackage{lscape}
\usepackage{subfig}
\usepackage{textcomp}
\usepackage{pstricks}
\usepackage[metapost]{mfpic}
\usepackage{fancybox}
\usepackage{slashed}
\usepackage[verbose]{wrapfig}
\usepackage{pifont}
\usepackage{keystroke}
\usepackage[title]{appendix}
\usepackage{enumitem}
\usepackage{array}
\usepackage{colortbl}
\usepackage{alltt}
\usepackage{boxedminipage}
\usepackage{tikz}
\usepackage{calc}
\usepackage{subfloat}
\usepackage{multirow}

\usepackage{cmll}
\usepackage{multicol}
\definecolor{color1}{RGB}{204,0,51}
\definecolor{color2}{RGB}{159,182,205}
\usepackage[open,openlevel=3]{bookmark}

\usepackage{verbatim}
\usepackage{pst-grad} 
\usepackage{pst-plot}
\usepackage{transparent}
\usepackage{color,soul}
\usepackage{cancel}
\usepackage{placeins}
\usepackage{mathtools}
\usepackage{rotating}
\usepackage[refpage]{nomencl}
\usepackage{hhline}
\usepackage{marginnote}
\usepackage{upgreek}
\usepackage{listings}
\usepackage{hyperref}
\usepackage{cleveref}
\usepackage{afterpage}
\usepackage{cite}
\usepackage{makecell}
\usepackage{accents}
\usepackage{breqn}
\usepackage{environ}
\usepackage{empheq}
\usepackage{bbding}
\usepackage{titlesec}
\usepackage[final]{showlabels}

\newcommand{\cc}[1]{\textcolor{ccolor}{c^{#1}}}
\definecolor{ccolor}{RGB}{255, 95, 31}
\definecolor{cdolor}{RGB}{255, 0, 0}

\newcommand{\tor}[0]{\mathfrak{t}}
\newcommand{\tore}[0]{\mathfrak{e}}

\newcommand{\tnew}[0]{\hat{\tau}}
\newcommand{\enew}[0]{\hat{e}}

\allowdisplaybreaks
\lstdefinestyle{customc}{
  belowcaptionskip=1\baselineskip,
  breaklines=true,
  frame=L,
  xleftmargin=\parindent,
  language=C,
  showstringspaces=false,
  basicstyle=\footnotesize\ttfamily,
  keywordstyle=\bfseries\color{green!40!black},
  commentstyle=\itshape\color{purple!40!black},
  identifierstyle=\color{blue},
  stringstyle=\color{red217!80!black},
}

\lstdefinestyle{customasm}{
  belowcaptionskip=1\baselineskip,
  frame=L,
  xleftmargin=\parindent,
  language=[x86masm]Assembler,
  basicstyle=\footnotesize\ttfamily,
  commentstyle=\itshape\color{purple!40!black},
}

\sethlcolor{yellow}

\hypersetup{
linkbordercolor={red}
}

\hypersetup{
    bookmarks=false,         
    unicode=false,          
    pdftoolbar=true,        
    pdfmenubar=true,        
    pdffitwindow=false,     
    pdfstartview={FitH},    
    pdftitle={Non-Relativistic Heterotic String Theory},    
    pdfauthor={Eric Bergshoeff, Luca Romano},     
    pdfsubject={},   
    pdfnewwindow=true,      
    colorlinks=false,       
    linkcolor=red,          
    citecolor=red,        
    filecolor=red,      
    urlcolor=red           
    linkbordercolor={red},
    citebordercolor={red},
    urlbordercolor={red},
    bookmarksopen={true}
}

\usepackage{listings} 

\makeatletter

\newcommand{\Rmnum}[1]{\expandafter\@slowromancap\romannumeral #1@}
\makeatother

\theoremstyle{definition}

\theoremstyle{remark}

\theoremstyle{proposition}

\usepackage{alphalph}

\makeatletter
\newalphalph{\aalphalph}[mult]{\alphalph@alph}{26}
\newcommand{\alphalphval}[1]{%
  \@ifundefined{c@#1}{
    \aalphalph{#1}
  }{%
    \aalphalph{\value{#1}}
  }
}
\makeatother

\usepackage{color}

\def\chapterautorefname~#1\null{Chap.~(#1)\null}
\def\sectionautorefname~#1\null{Sec.~(#1)\null}
\def\subsectionautorefname~#1\null{sub--Sec.~(#1)\null}
\def\figureautorefname~#1\null{Fig.~(#1)\null}
\def\tableautorefname~#1\null{Tab.~(#1)\null}
\def\equationautorefname~#1\null{~(#1)\null}

\def\equationautorefname~#1\null{eq.~(#1)\null}

\NewEnviron{multieq}[1][2]{

\begin{multicols}{#1}
\begin{subequations}
\setlength{\abovedisplayskip}{-11pt}
\allowdisplaybreaks
\begin{align}
\BODY
\end{align}
\end{subequations}
\end{multicols}
\setlength{\parindent}{0pt}
\noindent
}

\NewEnviron{multieqthree}[1][2]{

\begin{multicols}{#1}
\begin{subequations}
\setlength{\abovedisplayskip}{-16pt}
\allowdisplaybreaks
\begin{align}
\BODY
\end{align}
\end{subequations}
\end{multicols}
\setlength{\parindent}{0pt}
\noindent
}

\NewEnviron{multieqsep}[1][2]{

\begin{multicols}{#1}
\setlength{\columnseprule}{0.4pt}
\begin{subequations}
\setlength{\abovedisplayskip}{-12pt}
\allowdisplaybreaks
\begin{align}
\BODY
\end{align}
\end{subequations}
\end{multicols}
\setlength{\parindent}{0pt}
\noindent
}

\newcommand\multieqreference{}
\NewEnviron{multieqref}[2][2]{
\begin{multicols}{#1}
\def\multieqreference{#2}%
\begin{subequations}\label{\multieqreference}\
\setlength{\abovedisplayskip}{-15pt}
\allowdisplaybreaks
\begin{align}
\BODY
\end{align}
\end{subequations}
\end{multicols}
\setlength{\parindent}{0pt}
\noindent
}

\NewEnviron{multieqrefB}[2][2]{
\begin{subequations}
\begin{multicols}{#1}
\setlength{\abovedisplayskip}{-15pt}
\allowdisplaybreaks
\begin{align}
\BODY
\end{align}
\end{multicols}\label{#2}
\end{subequations}
\setlength{\parindent}{0pt}
\noindent
}

\DeclareMathAlphabet\mathbfcal{OMS}{cmsy}{b}{n}

\usepackage[yyyymmdd]{datetime}

\title{\huge\bf Non-Relativistic Heterotic String Theory}

\date{}

\begin{document}

\begin{flushright}
\small
October 30\textsuperscript{th}, 2023\\
\normalsize
\end{flushright}
{\let\newpage\relax\maketitle}
\maketitle
\def\equationautorefname~#1\null{(#1)\null}
\def\tableautorefname~#1\null{tab.~(#1)\null}
\def\sectionautorefname~#1\null{#1\null}

\vspace{0.8cm}

\begin{center}
\renewcommand{\thefootnote}{\alph{footnote}}
{\bf\large Eric~A.~Bergshoeff$^{~1}$}\footnote{Email: {\tt e.a.bergshoeff[at]rug.nl}  } {\bf and}
{\bf\large Luca~Romano$^{~2}$}\footnote{Email: {\tt lucaromano2607[at]gmail.com}}
\setcounter{footnote}{0}
\renewcommand{\thefootnote}{\arabic{footnote}}

\vspace{0.5cm}
${}^1${\it Van Swinderen Institute, University of Groningen\\
Nijenborgh 4, 9747 AG Groningen, The Netherlands}\\
\vskip .2truecm
${}^2${\it Departamento de Electromagnetismo y Electronica, Universidad de Murcia,\\ Campus de Espinardo, 30100 Murcia, Spain }\\

\vspace{1.8cm}


{\bf Abstract}
\end{center}
\begin{quotation}
{\small

In this work we consider heterotic-gravity as the low-energy approximation to heterotic string theory. We define a consistent non-relativistic limit of heterotic gravity that includes the Yang-Mills Chern-Simons term. We perform three tests on the heterotic  limit:  we use it to (i) derive the non-relativistic transformation rules, (ii) show the existence of a finite non-relativistic heterotic action and (iii) obtain the longitudinal  non-relativistic T-duality rules. We show that in all these cases the limit procedure is well defined,  free of divergences and leads to consistent non-trivial  results. We comment on the interpretation of the  T-duality rules in terms of a heterotic non-relativistic geometry underlying non-relativistic heterotic string theory.
}
\end{quotation}

\newpage

\tableofcontents

\section*{Introduction}\addcontentsline{toc}{section}{\protect\numberline{}Introduction}

An intriguing aspect of  non-relativistic string theory \cite{Gomis:2000bd,Danielsson:2000gi} is that under T-duality a spacelike longitudinal direction is mapped to a null direction underlying a Discrete Lightcone Quantization (DLCQ) of relativistic string theory \cite{Gomis:2000bd,Danielsson:2000gi,Bergshoeff:2018yvt}.
Non-relativistic string theory provides in this way a concrete realization of DLCQ string theory. This feature of non-relativistic T-duality has also been studied from a target space effective action point of view  both in the bosonic \cite{Harmark:2019upf,Bergshoeff:2021bmc}
 and the $\mathcal{N}=1$ supersymmetric case \cite{Bergshoeff:2021tfn}. In particular, it has been verified that  a longitudinal compactification of the (bosonic sector of) non-relativistic ten-dimensional $\mathcal{N}=1$ supergravity without Yang-Mills (YM) can be identified with a null reduction of the relativistic $\mathcal{N}=1$ supergravity theory \cite{Bergshoeff:2021bmc,Bergshoeff:2021tfn}.\\
\\
Sofar, the existence of similar features in heterotic string theory has been largely unexplored. From the sigma model point of view this has to do with the fact that a gauge-invariant heterotic sigma model   can only be defined at the level of the quantum effective action after the chiral sigma model anomaly, due to the heterotic fermions, have been taken into account \cite{Hull:1985jv}. It is in order to cancel this  anomaly that  the Kalb-Ramond (KR)  two-form field of the heterotic theory transforms non-trivially under Yang-Mills transformations that leads to a Yang-Mills Chern-Simons term in the curvature of the KR field. This property plays a key role in the famous Green-Schwarz anomaly cancellation that led to the first superstring revolution \cite{Green:1984sg}.\\
\\
From the target space point of view only very limited results have been obtained sofar. To be precise, it has been shown how a stringy limit of the 10D super Yang-Mills    system in flat spacetime, consistent with supersymmetry, can be obtained \cite{Bergshoeff:2021tfn}. A noteworthy feature of this stringy limit is that it requires that only one of the two longitudinal lightcone components of the YM vector field needs to be redefined with a contraction parameter. Including gravity, one expects that a new feature, motivated by the presence of the YM Chern-Simons term, might be that the redefinitions of the geometric fields, i.e.~the Zehnbein, the KR two-form and the dilaton, could involve the YM gauge field. Due to the complexity of the problem, it is non-trivial to guess what these new redefinitions should be.\\
\\
In this work we wish to report about a set of redefinitions that defines a consistent limit of (the bosonic sector of) 10D
heterotic supergravity. This theory is described by the following action:\,
\begin{align}\label{haction}
S_{\rm heterotic}&=\frac{1}{2}\int\, d^{10}x\, \sqrt{-g}\, e^{-2\Phi}\bigg[{\rm R}+4(\partial\Phi)^{2}-\frac{3}{4}\mathcal{H}^{2}+\frac{1}{2}F_{MN\, I }F^{MN\, I}\bigg]\,,
\end{align}
in terms of the basic relativistic fields
\begin{equation}
E_M{}^{\hat A}\,, B_{MN}\,, \Phi\,, V_M^I\,,
\end{equation}
where $E_M{}^{\hat A}$ is the Zehnbein that can be used to define a metric tensor
\begin{equation}
g_{MN} = E_M{}^{\hat A} E_N{}^{\hat B} \eta_{\hat A\hat B}\,,
\end{equation}
$\Phi$ is the dilaton, $B_{MN}$ is the KR two-form and $V_M^I$ is the YM gauge field, with curvatures given by
\begin{subequations}
\begin{align}
F_{MN}^{I}&=2\partial_{[M}V_{N]}^{I}-\sqrt{2}{\tt g}\, f_{KL}{}^{I}V_{M}^{K}V_{N}^{L}\,,\\
\mathcal{H}_{MNP}&=\partial_{[M}B_{NP]}-V_{[M}^{I}F_{NP]I}-\frac{\sqrt{2}}{3}{\tt g}\, f_{IJK}V_{[M}^{I}V_{N}^{J}V_{P]}^{K}\,,
\end{align}
\end{subequations}
where the last equation contains the Chern-Simons term.\\
\\
The Lagrangian \autoref{haction} is invariant under the following   Lorentz transformations with parameters $\Lambda^{\hat A}{}_{\hat B}$, one-form gauge transformations with parameters $\Lambda_M$ and YM gauge transformations with parameters $\Lambda^I$:
\begin{subequations}\label{transfrules}
\begin{align}
\delta E_{M}{}^{\hat{A}}&=\Lambda^{\hat{A}}{}_{\hat{B}}E_{M}{}^{\hat{B}}\,,\\
\delta B_{MN}&=2\partial_{[M}\Lambda_{N]}+2V_{[M}^{I}\partial_{N]}\Lambda_{I}\,,\\
\delta V^{I}_{M}&=\partial_{M}\Lambda^{I}+\sqrt{2}{\tt g}\, f_{JK}{}^{I}\Lambda^{J}V_{M}^{K}\,.
\end{align}
\end{subequations}
Apart from these symmetries the heterotic action is also invariant under
general coordinate transformations  under which the basic fields transform in the standard way. On top of this, the heterotic action is invariant under T-duality transformations, that can convenient be expressed in term of the usual metric, $g_{MN}$, and a generalized metric $G_{MN}$ given by \cite{Bergshoeff:1995cg}\,\footnote{This generalized metric was for the Abelian case noted in \cite{Giveon:1988tt, Shapere:1988zv}.}
\begin{equation}\label{effective}
G_{MN} = g_{MN} +B_{MN} - V_M^IV_{NI}\,.
\end{equation}
The challenge is to find a redefinition of the relativistic fields defining a non-relativistic limit such that after taking the limit the following results are non-trivial and  consistent and, in particular, that there are no divergences in
\begin{enumerate}
\item the transformation rules \autoref{transfrules}\,,
\item the heterotic action \autoref{haction}\,,
\item the T-duality rules \autoref{effective}\,.
\end{enumerate}

In general, a non-relativistic limit can be defined by first making an invertible redefinition of the  relativistic fields involving a contraction parameter for which we take the velocity of light  $\cc{}$.\,\footnote{More precisely, one should first introduce a dimensionless contraction parameter  $\lambda$ by redefining $\cc{} \to \lambda\, \cc{}$ and next take the limit that $\lambda \to \infty$.} The non-relativistic limit is then obtained by taking the limit that $\cc{}\to \infty$. Given the mixing of the geometric fields with the YM fields, it is a formidable task to find a proper redefinition that leads to a consistent limit by starting from a most general Ansatz. Instead, by focussing on
 the absence of divergences in the non-relativistic limit of the  T-duality rules, we found a set of redefinitions that, without derivation, will be the starting point of this work. We will use these redefinitions to show that indeed the above three properties hold. The details about the actual brute force derivation will be given in a larger companion paper. Given these results, it seems plausible that the set of redefinitions that defines our non-relativistic limit is unique up to redefinitions of the fields.\\
\\
One interesting result  is that the non-relativistic T-duality rules we find do not convert a spatial longitudinal direction into a null direction with respect to the dual metric tensor $\tilde g_{MN}$ but into a null direction with respect to the dual  generalized metric tensor $\tilde G_{MN}$. This should have implications for the relation between non-relativistic heterotic string theory and the DLCQ of heterotic string theory.\\
\\
This work is organized as follows. In the next section we present our redefinitions  defining the non-relativistic limit. In the following sections we will show that this limit satisfies the three properties mentioned above. More precisely, first in section \autoref{sec:NRS} will  derive the finite transformation rules. Next, in section \autoref{sec:NRA} we will show that all divergences cancel when taking the non-relativistic limit of the heterotic action. Finally,
in section \autoref{sec:NRT} we will derive the non-relativistic T-duality rules.
We give our conclusions and outlook in a separate section and have added an appendix explaining our notation and conventions.

\section{Defining the Limit} \label{sec:ansatz}
In this section we present our starting point, i.e.~the redefinition for all relativistic fields of the heterotic theory.  These redefinitions define the non-relativistic limit that, in the next sections, we will apply to the transformation rules \autoref{transfrules} of the relativistic fields, the heterotic action \autoref{haction} and the heterotic T-duality rules involving the generalized metric \autoref{effective}.
To be concrete, decomposing the flat index $\hat A$ as $\hat A = (A,a) = (-,+,a)$, we propose the following invertible redefinitions:\,\footnote{We remind that our notation and conventions can be found in \autoref{sec:Notation}.}
\begin{subequations}\label{eq:Expansion}
\begin{align}
E_{M}{}^{-}&=\cc{}\, \tau_{M}{}^{-}\,,\\
E_{M}{}^{+}&=-\cc{3}\frac{v_{-}^{2}}{2}\, \tau_{M}{}^{-}+\cc{}\, \tau_{M}{}^{+}\,,\\
E_{M}{}^{a}&=e_{M}{}^{a}\,,\\
B_{MN}&=\cc{2}\tau_{M}{}^{A}\tau_{N}{}^{B}\epsilon_{AB}(1+v_{+-})
+2\cc{2}\tau_{[M}{}^{-}e_{N]}{}^{a}v_{-a}+b_{MN}\,,\\
\Phi &= \phi + \log \cc{}\,,\\
V_{M}^{I}&=\cc{2}\tau_{M}{}^{-}v_{-}^{I}+\tau_{M}{}^{+}v_{+}^{I}+e_{M}{}^{a}v_{a}^{I}\,.
\end{align}
\end{subequations}
 The fields
\begin{equation}
\tau_{M}{}^{A},\, e_{M}{}^{a},\,  b_{MN}\,, \phi\,, v_{\pm}^{I},\, v_{a}^{I}
\end{equation}
 will be the non-relativistic fields after sending $\cc{}$ to infinity.
The  redefinitions \autoref{eq:Expansion} imply the following redefinitions of the metric $g_{MN}$ and the generalized metric $G_{MN}$:
\begin{subequations}\label{eq:Expansion2}
\begin{align}
g_{MN}&=\cc{4}\tau_{M}{}^{-}\tau_{N}{}^{-}v_{-}^{2}+\cc{2}\tau_{M}{}^{A}\tau_{N}{}^{B}\eta_{AB}+h_{MN}\label{eq:ExpansionMetric}\,,\\
G_{MN}&=-2\cc{2}\Big[\tau_{M}{}^{+}\tau_{N}{}^{-}(1+v_{+-})+e_{M}{}^{a}\tau_{N}{}^{-}v_{-a}\Big]+\nonumber\\
&\hskip .3truecm +h_{MN}+b_{MN}-(\tau_{M}{}^{+}v_{+}^{I}+e_{M}{}^{a}v_{a}^{I})(\tau_{N}{}^{+}v_{+I}+e_{N}{}^{b}v_{bI})\,.\label{eq:Expensioneffectivemetric}
\end{align}
\end{subequations}
Here and above,  we have used the abbreviations
\begin{multieq}[2]
h_{MN}&=e_M{}^a e_N{}^b \delta_{ab}\,,\\
v_{-a}&= v_-^I v_{aI}\,,\\
v_{+-}&= v_+^I v_{-I}\,,\\
v_-^2 &= v_-^I v_{-I}\,.
\end{multieq}

We notice that, as expected,  both the expansion of the longitudinal vielbein field component $E_M{}^+$  as well as the expansion of the KR 2-form field $B_{MN}$ in \autoref{eq:Expansion} contain extra terms proportional to the YM vector fields that are absent in
 the usual redefinitions leading to Newton-Cartan geometry. We furthermore point out that the
  $\cc{2}$ term in the redefinition of $B_{MN}$ that does not involve the YM vector fields  differs by a sign from that appearing in \cite{Bergshoeff:2021tfn}. Setting to zero the vector field we recover the ansatz of  \cite{Bergshoeff:2021tfn} only  up to a redefinition. Note that the sign of the $\cc{2}$ term in the  redefinition of $B_{MN}$ not containing the YM vector field is not fixed  by requiring cancellation of the divergences in the action, since it appears via a quadratic term in that action, neither can it be fixed  by considering the limit of  the T-duality rules. Only when taking into account the YM vector fields, the T-duality rules become sensitive to that sign and we need to fix it according to \autoref{eq:Expansion}.\\
\\
In the next three sections we will test the  redefinitions \autoref{eq:Expansion} as described in the introduction.

\section{The Non-Relativistic Symmetries}\label{sec:NRS}
In this section we perform the limit of the transformation rules governing heterotic gravity, using the redefinitions \autoref{eq:Expansion}. We do not include in our discussion the dilaton, due to the triviality of its transformation rules, and omit diffeomorphisms, since they are unmodified by the limit.  Our starting point is the set of relativistic transformations \autoref{transfrules}.\\
\\
When redefining the Lorentz parameters $\Lambda_{\hat{A}\hat{B}}$, we will decompose them into transverse Lorentz parameters $\Lambda_{ab}$, longitudinal Lorentz parameters $\Lambda$ and Lorentz boost parameters $\Lambda_{Aa}$.   The structure of the longitudinal Lorentz, transverse Lorentz and the Abelian 1-form gauge transformations require a trivial redefinition of the parameters. Since these transformations look the same before and after taking the limit we will not give further details  here. Rather, we wish to discuss the non-trivial limit of the boost and non-Abelian gauge transformations. Note that all the transformation rules given below reduce to those of standard string Newton-Cartan geometry upon setting the YM fields to zero.\\
\\
We first discuss the boost transformations. Avoiding divergences in the limit requires, on top of the field redefinitions \autoref{eq:Expansion}, the following non-trivial redefinition of the boost parameters:

\begin{multieq}[2]
\Lambda_{a+}&=\frac{1}{\cc{}}\lambda_{a+}(1+v_{+-})\,,\\
\Lambda_{a-}&=\frac{1}{\cc{}}\lambda_{a-}+\frac{\cc{}}{2}\lambda_{a+}v_{-}^{2}(1+v_{+-})\,.
\end{multieq}
Using these redefinitions of the parameters together with the redefinitions \autoref{eq:Expansion} of the fields, we find the following finite boost transformations:
\begin{subequations}
\begin{align}
\delta\,  \tau_{M}{}^{-}&=0\,,\\
\delta\,  \tau_{M}{}^{+}&=\frac{\lambda_{-}{}^{a}v_{a-}}{1+v_{+-}}\tau_{M}{}^{-}-e_{M}{}^{a}\lambda_{+a}v_{-}^{2}(1+v_{+-})\,,\\
\delta\,  e_{M}{}^{a}&=-\lambda_{-}{}^{a}\tau_{M}{}^{-}-\lambda_{+}{}^{a}\tau_{M}{}^{+}(1+v_{+-})\,,\\
\delta\,  v_{+}^{I}&=\lambda_{+}{}^{a}(1+v_{+-})v_{a}^{I}\,,\\
\delta\,  v_{-}^{I}&=0\,,\\
\delta\,  v_{a}^{I}&=\lambda_{+a}(1+v_{+-})(v_{-}^{I}+v_{-}^{2}v_{+}^{I})\,,\\
\delta\,  b_{MN}&=-\tau_{M}{}^{A}\tau_{N}{}^{B}\epsilon_{AB}\lambda_{-}{}^{a}\bigg(v_{a+}-\frac{v_{+}^{2}}{1+v_{+-}}v_{a-}\bigg)+2\tau_{[M}{}^{-}e_{N]}{}^{a}\lambda_{-}{}^{b}\bigg[\delta_{ab}-\bigg(v_{ab}-\frac{v_{a+}v_{b-}}{1+v_{+-}}\bigg)\bigg]+\nonumber\\
&+2\lambda_{+a}e_{[M}{}^{a}\Big[\tau_{N]}{}^{+}(1+v_{+-})+e_{N]}{}^{b}v_{b-}\Big](1+v_{+-})\,.
\end{align}
\end{subequations}
\setlength{\parindent}{0pt}

We next consider the non-Abelian gauge transformation. After making a trivial redefinition of the gauge parameter
\begin{align}
\Lambda^{I}&=\sigma^{I}\,,
\end{align}
and sending $\cc{}\rightarrow \infty$, we obtain the following finite non-Abelian gauge transformation rules:
\begin{subequations}
\begin{multicols}{2}
\setlength{\abovedisplayskip}{-11pt}
\allowdisplaybreaks
\begin{align}
\delta\,  \tau_{M}{}^{-}&=0\,,\\
\delta\,  e_{M}{}^{a}&=0\,,\\
\delta\,  \tau_{M}{}^{+}&=\frac{v_{-I}\partial_{-}\sigma^{I}}{1+v_{+-}}\tau_{M}{}^{-}\,,\\
\delta\,  v_{+}^{I}&=\partial_{+}\sigma^{I}+\sqrt{2}{\tt g} f_{JK}{}^{I}\sigma^{J}v_{+}^{K}\,,\\
\delta\,  v_{-}^{I}&=\sqrt{2}{\tt g} f_{JK}{}^{I}\sigma^{J}v_{-}^{K}\,,\\
\delta\,  v_{a}^{I}&=\partial_{a}\sigma^{I}+\sqrt{2}{\tt g} f_{JK}{}^{I}\sigma^{J}v_{a}^{K}\,,
\end{align}
\end{multicols}
\vspace{-0.6cm}
\begin{align}
\delta\,  b_{MN}&=2\tau_{[M}{}^{+}e_{N]}{}^{a}\Big(v_{+I}\partial_{a}\sigma^{I}-v_{aI}\partial_{+}\sigma^{I}\Big)-2e_{[M}{}^{a}e_{N]}{}^{b}v_{a}^{I}\partial_{b}\sigma_{I}+\nonumber\\
&+2\bigg[\tau_{[M}{}^{-}\tau_{N]}{}^{+}\bigg(-2v_{+I}+\frac{v_{+}^{2}}{1+v_{+-}}v_{-I}\bigg)+\tau_{[M}{}^{-}e_{N]}{}^{a}\bigg(-2v_{aI}+\frac{v_{a+}}{1+v_{+-}}v_{-I}\bigg)\bigg]\partial_{-}\sigma^{I}\,.
\end{align}
\end{subequations}
\setlength{\parindent}{0pt}
Here, the derivatives $\partial_\pm$ and $\partial_a$ are defined by $\partial_\pm = \tau_\pm{}^M\partial_M$ and $\partial_a = e_a{}^M\partial_M$, respectively.\\

We have verified  that the boost and non-Abelian gauge transformations, together with the other non-Lorentzian gauge transformations define a
Lie algebra. The corresponding non-trivial commutation relations will be given in the longer companion paper.

\section{The Divergences of the Heterotic Lagrangian}\label{sec:NRA}

We now perform our second test and show that upon substituting the redefinitions  \autoref{eq:Expansion}
into the heterotic action \autoref{haction}  all divergent terms cancel. More precisely, we find that
the redefinitions
\autoref{eq:Expansion} induce the following expansion of the Ricci scalar ${\rm R}$, the 2-form curvature $\mathcal{H}_{MNP}$ and the YM curvature $F_{MN}^{I}$
\begin{subequations}
\begin{align}
{\rm R}&=\cc{4}\accentset{(4)}{{\rm R}}+\cc{2}\accentset{(2)}{{\rm R}}+\accentset{(0)}{{\rm R}}+\frac{1}{\cc{2}}\accentset{(-2)}{{\rm R}}\,,\\
\mathcal{H}_{MNP}&=\cc{4}\accentset{(4)}{\mathcal{H}}_{MNP}+
\cc{2}\accentset{(2)}{\mathcal{H}}_{MNP}+\accentset{(0)}{\mathcal{H}}_{MNP}\,,\\
F_{MN}^{I}&=\cc{2}\accentset{(2)}{F}^{I}_{MN}+\accentset{(0)}{F}^{I}_{MN}\,,
\end{align}
\end{subequations}
with the different terms given by\,\footnote{for the definitions of $\tor_{MN}{}^A$ and $\tore_{MN}{}^a$, see the appendix.}
\begin{subequations}
\begin{align}
\accentset{(4)}{{\rm R}}&=\frac{1}{2}\tor^{-a-}\tor^{-}{}_{a}{}^{-}v_{-}^{4}-\frac{1}{4}\tor_{ab}{}^{-}\tor^{ab-}v_{-}^{2}\,,\\
\nonumber\\
\accentset{(2)}{{\rm R}}&=\tor^{-a-}\partial_{a}v_{-}^{2}+\frac{1}{2}\tor_{ab}{}^{+}\tor^{ab-}+(\tor^{a+-}\tor_{a}{}^{--}-\tor^{a-+}\tor_{a}{}^{--})v_{-}^{2}\,,\\
\nonumber\\
\accentset{(2)}{F}_{M N}{}^{I}&= -2\tau_{[M}{}^{-} \partial_{N]}v_{-}^{I}+\tor_{M N}{}^{-} v_{-}^{I}+\nonumber\\
&-2\sqrt{2} {\tt g} \, \tau_{[M}{}^{-} e_{N]}\,^{a} f^{IJK} v_{- J} v_{a K}+2\sqrt{2} {\tt g} \tau_{[M}{}^{+} \tau_{N]}{}^{-} f^{IJK} v_{- J} v_{+K}\,,\\
\nonumber\\
\accentset{(0)}{F}_{M N}{}^{I} &= {\tore}_{M N}\,^{a} v_{a}^{I}+\tor_{M N}{}^{+} v_{+}^{I}-2\tau_{[M}{}^{+} \partial_{N]}v_{+}^{I}-2e_{[M}\,^{a} \partial_{N]}v_{a}^{I}+\nonumber\\
&-2\sqrt{2} {\tt g}\, \tau_{[M}{}^{+} e_{N]}\,^{a} f^{IJK} v_{+ J} v_{a K}-\sqrt{2} {\tt g} e_{M}\,^{a} e_{N}{}^{b} f^{IJK} v_{a J} v_{b K}\,,\\
\nonumber\\
\accentset{(4)}{\mathcal{H}}_{MNP}&=-\tau_{[M}{}^{-}\tor_{NP]}{}^{-}v_{-}^{2}\,,\\
\nonumber\\
\accentset{(2)}{\mathcal{H}}_{MNP}&=-\tau_{[M}{}^{A}\tor_{NP]}{}^{B}\epsilon_{AB}-2\tau_{[M}{}^{-}\tore_{NP]}{}^{a}v_{a-}-
2\tau_{[M}{}^{-}\tor_{NP]}{}^{+}v_{+-}+\nonumber\\
&+4v_{-I}\tau_{[M}{}^{-}e_{N}{}^{a}\partial_{P]}v_{a}^{I}-4v_{-I}\tau_{[M}{}^{+}\tau_{N}{}^{-}\partial_{P]}v_{+}^{I}+\nonumber\\
&+2\sqrt{2}{\tt g}\, \tau_{[M}{}^{-}e_{N}{}^{a}e_{P]}{}^{b}f_{IJK}v_{-}^{I}v_{a}^{J}v_{b}^{K}-4\sqrt{2}{\tt g}\,\tau_{[M}{}^{+}\tau_{N}{}^{-}e_{P]}{}^{a}f_{IJK}v_{-}^{I}v_{+}^{J}v_{a}^{K}\,,\\
\nonumber\\
\accentset{(0)}{\mathcal{H}}_{MNP}&=h_{MNP}-\tau_{[M}{}^{+}\tor_{NP]}{}^{+}v_{+}^{2}-\tau_{[M}{}^{+}\tore_{NP]}{}^{a}v_{+a}-e_{[M}{}^{a}\tore_{NP]}{}^{b}v_{ab}-e_{[M}{}^{a}\tor_{NP]}{}^{+}v_{a+}+\nonumber\\
&-2v_{aI}\tau_{[M}{}^{+}e_{N}{}^{a}\partial_{P]}v_{+}^{I}
+2v_{+I}\tau_{[M}{}^{+}e_{N}{}^{a}\partial_{P]}v_{a}^{I}
+2v_{aI}e_{[M}{}^{a}e_{N}{}^{b}\partial_{P]}v_{b}^{I}+\nonumber\\
&+4\sqrt{2}{\tt g}\, \tau_{[M}{}^{+}e_{N}{}^{a}e_{P]}{}^{b}f_{IJK}v_{-}^{I}v_{a}^{J}v_{b}^{K}\,.
\end{align}
\end{subequations}
\\
There are potentially  diverging terms in the action  at orders $\cc{4}$ and $\cc{2}$. We find that the sum of the three diverging contributions coming from the Ricci scalar, the 3-form field strength and the 2-form YM field strength precisely cancel both at order $\cc{4}$ and $\cc{2}$. In particular at order $\cc{4}$ the contribution to the action of the 2-form and 3-form field strengths are
\begin{subequations}
\begin{align}
F_{MN\, I}F^{MN\, I}&=\cc{4}\Big(\tor_{ab}{}^{-}\tor^{ab-}v_{-}^{2}-2\tor^{a--}\tor_{a}{}^{--}v_{-}^{4}\Big)+\mathcal{O}(\cc{2})\,,\\
\mathcal{H}_{MNP}\mathcal{H}^{MNP}&=\frac{1}{3}\cc{4}\Big(\tor_{ab}{}^{-}\tor^{ab-}v_{-}^{2}-2\tor^{a--}\tor_{a}{}^{--}v_{-}^{4}\Big)+\mathcal{O}(\cc{2})\,.
\end{align}
\end{subequations}
It is immediate to recognize that
\begin{multieq}[2]
F_{MN\, I}F^{MN\, I}&=-4\cc{4}\accentset{(4)}{{\rm R}}+\mathcal{O}(\cc{2})\,,\\
\mathcal{H}_{MNP}\mathcal{H}^{MNP}&=-\frac{4}{3}\cc{4}\accentset{(4)}{{\rm R}}+\mathcal{O}(\cc{2})\,,
\end{multieq}
leading to a full cancellation in \autoref{haction}. At order $\cc{2}$ the cancellation mechanism is more involved due to a more intricate interplay between the three terms. The computations have been carried out with the help of Cadabra \cite{Peeters2018, Peeters:2007wn, PEETERS2007550}. This shows  the existence of a finite limit of the heterotic action. Due to its complexity, we refrain from giving the explicit expression of this action. Obtaining a managable expression probably requires a proper understanding of the underlying heterotic geometry.

\section{Non-Relativistic Longitudinal T-duality}\label{sec:NRT}

Finally, we test our set of redefinitions  \autoref{eq:Expansion} against the  heterotic T-duality rules. Using the generalized metric $G_{MN}$ given in eq.~\autoref{effective} the relativistic heterotic  T-duality rules  are given by (see eq.~(39) of \cite{Bergshoeff:1995cg})
\begin{multicols}{2}
\begin{subequations}\label{eq:T-duality0b}
\setlength{\abovedisplayskip}{-15pt}
\allowdisplaybreaks
\begin{align}
\tilde{g}_{\mu\nu}&=g_{\mu\nu}+\frac{g_{xx}G_{x\mu}G_{x\nu}-2G_{xx}G_{x(\mu}g_{\nu)x}}{G_{xx}^2}\ ,\label{eq:gmunu0b}\\
\tilde{B}_{\mu\nu}&=B_{\mu\nu}{+}\frac{G_{x[\mu}G_{\nu]x}}{G_{xx}}\label{eq:Bmunu0b}\ ,\\
\tilde{g}_{x\mu}&=-\frac{g_{x\mu}}{G_{xx}}+\frac{g_{xx}G_{x\mu}}{G_{xx}^2}\ ,\\
\tilde{B}_{x\mu}&=\frac{G_{x\mu}-B_{x\mu}}{G_{xx}}\ ,\\
\tilde{g}_{xx}&=\frac{g_{xx}}{G_{xx}^{2}}\ ,\\
\tilde{\Phi}&=\Phi-\frac{1}{2}\log|G_{xx}|\ ,\label{eq:T-dualityDilaton0b}\\
\tilde{V}_{x}^{I}&=-\frac{V_{x}^{I}}{G_{xx}}\ ,\\
\tilde{V}_{\mu}^{I}&=V_{\mu}^{I}-\frac{V_{x}^{I}G_{x\mu}}{G_{xx}}\ ,
\end{align}
\end{subequations}
\end{multicols}
\setlength{\parindent}{0pt}
where we have decomposed  $M=\{\mu, x\}$, $x$ being the isometry direction.

We are interested in obtaining a non-relativistic spatial longitudinal T-duality. Decomposing the flat index ${\hat A}$ as ${\hat A} = (0,1;a)$, this  requires that we take the flat components of the Killing vector defining the isometry direction only non-zero in the 1 direction. To do this in an invariant way requires the following redefinition of the  Vierbeine and their inverses:
\begin{multieq}[2]
\tnew_{M}{}^{-}&=\tau_{M}{}^{-}\,,\\
\tnew_{M}{}^{+}&=\tau_{M}{}^{+} (1+v_{+-})+e_{M}{}^{a}v_{a-}\,, \phantom{\frac{1}{v_{+-}}}\\
\enew_{M}{}^{a}&=e_{M}{}^{a}\,, \phantom{\frac{1}{v_{+-}}}\\
\tnew^{M}{}_{-}&=\tau^{M}{}_{-}\,,\\
\tnew^{M}{}_{+}&=\frac{1}{1+v_{+-}}\tau^{M}{}_{+}\,,\\
\enew^{M}{}_{a}&=e^{M}{}_{a}-\frac{v_{a-}}{1+v_{+-}}\tau^{M}{}_{+}\,.
\end{multieq}
The hatted Vierbeine and their inverses satisfy the usual inverse relations of Newton-Cartan Geometry. They transform as follows  under YM and boost transformations:
\begin{multieq}[2]
\delta\, \tnew_{M}{}^{-}&=0\,,\\
\delta\, \tnew_{M}{}^{+}&=v_{-}^{I}\partial_{M}\sigma_{I}\,,\\
\delta\, \enew_{M}{}^{a}&=-\lambda_{A}{}^{a}\tnew_{M}{}^{A}+\lambda_{+}{}^{a}\enew_{M}{}^{b}v_{b-}\,,\\
\delta\, \tnew^{M}{}_{-}&=\lambda_{-}{}^{a}\enew^{M}{}_{a}-
v_{-I}\partial_{-}\sigma^{I}\tnew^{M}{}_{+}\,,\\
\delta\, \tnew^{M}{}_{+}&=\lambda_{+}{}^{a}\enew^{M}{}_{a}-
v_{-I}\tnew^{N}{}_{+}\partial_{N}\sigma^{I}\tnew^{M}{}_{+}\,,\\
\delta\, \enew^{M}{}_{a}&=-\lambda_{+}{}^{b}v_{-a}\enew^{M}{}_{b}-
v_{-I}\enew^{N}{}_{a}\partial_{N}\sigma^{I}\tnew^{M}{}_{+}\,.
\end{multieq}
The longitudinal spatial  isometry direction can then be defined by imposing the following conditions:
\begin{multicols}{3}
\begin{subequations}
\setlength{\abovedisplayskip}{-13pt}
\allowdisplaybreaks
\begin{align}
\tnew_{x}{}^{0}&=0\,,\\
\tnew_{x}{}^{1}&\neq 0\,,\\
\nonumber\\
\tnew_{x}{}^{a}&=0\,.
\end{align}
\end{subequations}
\end{multicols}
\setlength{\parindent}{0pt}
\noindent

These conditions break  half of the boost transformations, i.e.~$\lambda_{1a}=0$.\\
\\

Performing the limit on the above heterotic T-duality rules and defining the auxiliary expressions $\tnew_{MN}$ and $\ell_\mu$ as follows
\begin{subequations}
\begin{align}
\tnew_{MN}&=\tnew_{M}{}^{A}\tnew_{N}{}^{B}\eta_{AB}\,,\\
\ell_{\mu}&= (\tau_{x}{}^{+}v_{+I}+e_{x}{}^{a}v_{aI})\bigg[\bigg(\tau_{\mu}{}^{+}-\frac{\tau_{x}{}^{+}}{\tau_{x}{}^{-}}\tau_{\mu}{}^{-}\bigg)v_{+}^{I}+\bigg(e_{\mu}{}^{b}-\frac{e_{x}{}^{b}}{\tau_{x}{}^{-}}\tau_{\mu}{}^{-}\bigg) v_{b}^{I}\bigg]+h_{xx}\frac{\tau_{\mu}{}^{-}}{\tau_{x}{}^{-}}-h_{x\mu}-b_{x\mu}\,,
\end{align}
\end{subequations}
we find that all divergences cancel and that the remaining non-relativistic heterotic T-dualities are given by
\begin{subequations}\label{eq:NRTDuality1b}
\begin{align}
\tilde{\Phi}&=\phi-\frac{1}{2}\log|\tnew_{xx}|\,,\\
\nonumber\\
\tilde{g}_{xx}&=\frac{v_{-}^{2}(\tnew_{x}{}^{-})^{2}}{\tnew_{xx}^{2}}\,,\\
\nonumber\\
\tilde{V}_{x}^{I}&=-\frac{\tnew_{x}{}^{-}v_{-}^{I}}{\tnew_{xx}}\,,\\
\nonumber\\
\tilde{V}_{\mu}^{I}&=
v_{+}^{I}\bigg(\tau_{\mu}{}^{+}-\frac{\tau_{x}{}^{+}}{\tnew_{x}{}^{-}}\tnew_{\mu}{}^{-}\bigg)+v_{a}^{I}\bigg(e_{\mu}{}^{a}-\frac{e_{x}{}^{a}}{\tnew_{x}{}^{-}}\tnew_{\mu}{}^{-}\bigg)+\frac{\tau_{x}{}^{-}v_{-}^{I}}{\tnew_{xx}}\ell_{\mu}\,,\\
\nonumber\\
\tilde{B}_{x\mu}&=\frac{\tnew_{x\mu}}{\tnew_{xx}}\,,\\
\nonumber\\
\tilde{B}_{\mu\nu}&=b_{\mu\nu}+\frac{2}{\tnew_{x}{}^{-}}\tnew_{[\mu}{}^{-}b_{\nu]x}-\frac{2}{\tnew_{xx}}\ell_{[\mu}\tnew_{\nu]}{}^{A}\tnew_{x}{}^{B}\epsilon_{AB}\,,\\
\nonumber\\
\tilde{g}_{\mu x}&=\frac{\tau_{x}{}^{A}\tau_{\mu}{}^{B}\epsilon_{AB}}{\tnew_{xx}}-\frac{v_{-}^{2}(\tnew_{x}{}^{-})^{2}}{\tnew_{xx}^{2}}\ell_{\mu}\,,\\
\nonumber\\
\tilde{g}_{\mu\nu}&=\bigg(e_{\mu a}-\frac{\tnew_{\mu}{}^{-}}{\tnew_{x}{}^{-}}e_{x a}\bigg)\bigg(e_{\nu}{}^{a}-\frac{\tnew_{\nu}{}^{-}}{\tnew_{x}{}^{-}}e_{x}{}^{a}\bigg)+\frac{v_{-}^{2}(\tnew_{x}{}^{-})^{2}}{\tnew_{xx}^{2}}\ell_{\mu}\ell_{\nu}+\frac{2}{\tnew_{xx}}\ell_{(\mu}\tau_{\nu)}{}^{A}\tau_{x}{}^{B}\epsilon_{AB}\,.
\end{align}
\end{subequations}

We verified that, after setting the YM gauge fields equal to zero, we precisely reproduce the non-relativistic T-duality rules given in  equations (3.12) of \cite{Bergshoeff:2018yvt}  up to the redefinition that exchanges the two light-like directions (provided that the epsilon tensors are defined in the same way).\\
\\
It is instructive to also give the following T-duality rules of the generalized metric that explicitly show the presence of a null-isometry direction:
\begin{subequations}
\begin{multicols}{3}
\setlength{\abovedisplayskip}{-11pt}
\allowdisplaybreaks
\begin{align}
\tilde{G}_{xx}&=0\,,\\
\tilde{G}_{x\mu}&=\frac{\tnew_{\mu}{}^{-}}{\tnew_{x}{}^{-}}\,,\\
\tilde{G}_{x\mu}&=\frac{\tnew_{\mu}{}^{+}}{\tnew_{x}{}^{+}}\,,
\end{align}
\end{multicols}
\vspace{-0.6cm}
\begin{align}
\tilde{G}_{\mu\nu}&=h_{\mu\nu}+b_{\mu\nu}-(\tau_{\mu}{}^{+}v_{+I}+e_{\mu}{}^{a}v_{aI})(\tau_{\nu}{}^{+}v_{+}^{I}+e_{\nu}{}^{b}v_{b}^{I})\, +\nonumber\\
&+\frac{2}{\tnew_{xx}}\Big[\tnew_{\mu}{}^{-}\tnew_{\nu}{}^{+}-2\tnew_{\mu}{}^{+}\tnew_{\nu}{}^{-}\Big(h_{xx}-(\tau_{x}{}^{+}v_{+I}+e_{x}{}^{a}v_{aI})(\tau_{x}{}^{+}v_{+}	^{I}+e_{x}{}^{b}v_{b}^{I})\Big)\Big]\,.
\end{align}
\end{subequations}
\setlength{\parindent}{0pt}
We note that in the absence of Yang-Mills, one can also define a non-relativistic transverse and lightlike T-duality \cite{Bergshoeff:2019pij}. However, the way they are derived in \cite{Bergshoeff:2019pij} makes use of a sigma model description which is non-trivial in the heterotic case. Therefore, we do not consider these two cases further in this work.

\section*{Conclusions and Outlook}\addcontentsline{toc}{section}{\protect\numberline{}Conclusions and Outlook}

In this work we showed the existence of a consistent non-relativistic limit of heterotic gravity leading to  finite transformation rules,  a finite heterotic action and finite longitudinal T-duality rules.  Our results contain two noteworthy features. First of all, we find that under heterotic T-duality, a spatial longitudinal direction is mapped to a null isometry direction with respect to the generalized metric \autoref{effective}. It would be interesting to see what the role of this generalized metric is at the level of the sigma model description of the heterotic string and, in particular, whether the non-relativistic heterotic T-duality rules can be derived from a worldsheet duality transformation in the same way as this happens for the bosonic string without YM. We expect that the heterotic T-duality rules can be considerably simplified by formulating it as a $\mathbb{Z}_2$ transformation in a lower-dimensional theory. Such a reduction would involve a constrained spatial reduction expressing the fact that there is a null direction with respect to the generalized metric but not with respect to the metric. At a more fundamental level, it would be interesting to see what the impact of this new feature is on the DLCQ of the heterotic string and whether the YM transformations have some non-relativistic  M-theory interpretation like in the relativistic case \cite{Horava:1995qa}\,\footnote{We thank Ziqi Yan for commenting on this.}. For more literature on this, see
\cite{Danielsson:1996es, Kachru:1996nd, Motl:1997tb,Ebert:2023hba}.\\
\\
A second noteworthy feature is that, whereas in the relativistic case only the KR  2-form field transforms under YM transformations (apart from the YM gauge fields themselves), we find that in the non-relativistic case there is also a geometric field that transforms under YM transformations.
This makes the formulation of a geometric structure underlying the non-relativistic heterotic string theory non-trivial. Understanding the underlying geometry is crucial if one wants to present the finite part of the non-relativistic heterotic action in a simple way. A non-trivial feature is that the boost and YM transformations do not commute. We expect that therefore  the YM transformations are part of the structure group. One way to obtain more information about the geometric structure, is to consider the supersymmetric case. It is well-known that, in order to keep the supersymmetry rules finite, one needs to impose constraints on the intrinsic torsion tensors thereby restricting the geometry. Similarly, a better geometric understanding of the non-relativistic transformation rules might be obtained by looking for polynomial realizations in the same way as this has been done in the case of the non-relativistic $SL(2,\mathbb{R})$ symmetry underlying non-relativistic IIB supergravity \cite{Bergshoeff:2023ogz}. This work should serve as a starting point to understand the correct geometry underlying non-relativistic heterotic string theory. We hope to report about this in our longer companion paper where also more details of our calculations will be given. \\
\\
We notice that a different kind of duality has been observed for Yang-Mills in flat spacetime. One can show that taking a stringy limit of a Yang-Mills theory \cite{Bergshoeff:2015sic,Bergshoeff:2021tfn} followed by a longitudinal spatial reduction leads to the same model as a null reduction in flat spacetime of the same Yang-Mills theory\cite{Festuccia:2016caf}. It would be interesting to see whether this duality in one way or another can be viewed as  a special flat spacetime limit of the T-duality considered in this paper. \\
\\
Finally, a very practical application of the non-relativistic heterotic T-duality rules we derived is that they can be used as a solution generating transformation to obtain interesting solutions of non-relativistic heterotic string theory. We hope to come back to this interesting application soon.

\section*{Acknowledgements}
\noindent We thank Johannes Lahnsteiner, Jan Rosseel and Ceyda \c Sim\c sek for their involvement in the early explorations of this project. We furthermore wish to thank Johannes Lahnsteiner, Jan Rosseel  and Ziqi Yan for important comments on an earlier version of this work. The work of LR has been supported by Next Generation EU through the Maria Zambrano grant from the Spanish Ministry of Universities under the Plan de Recuperacion, Transformacion y Resiliencia. LR thanks the University of Groningen for its hospitality.

\appendix
\section{Notation and Convention}\label{sec:Notation}
We have worked in a curved ten-dimensional spacetime, adopting the following index notation
\FloatBarrier
\begin{table*}[!ht]
\centering
\renewcommand{\arraystretch}{1.5}
\begin{center}
\begin{tabular}{lcl}
\bf {Index}&&\bf{Definition \& Values}\\
$M,N,P,...$&& 10D Curved Index ($M=\{x,\mu\}$)\,,\\
$\mu,\nu,\rho,...$&& 9D Curved Index\,,\\
$x$&&Spatial or Null Isometry Direction\,, \\
$\hat{A},\hat{B},\hat{C},...$&&10D Flat Index ($\hat{A}=\{A,a\}$)\,,\\
$A,B,C...$&&Longitudinal Flat Index $A=0,1$\,,\\
$a ,b ,c ...$&&Transverse Flat Index $a=2,...,9$\,,\\
$I ,J ,K, L$&& Yang-Mills Gauge Algebra Adjoint Irrep. Index\,.
\end{tabular}
\end{center}
\end{table*}
\FloatBarrier
We denote with $g_{MN}$ and $\eta_{\hat{A}\hat{B}}$ the ten-dimensional curved and flat Minkowski metrics respectively, adopting for the latter the mostly plus signature. The ten-dimensional flat metric decomposes in a longitudinal metric $\eta_{AB}={\rm diag}(-1,+1)$ and a transverse Euclidean metric, $\delta_{ab}$. The longitudinal Levi-Civita tensor is chosen such that $\epsilon_{01}=+1$.  We denote with $f_{JK}{}^{I}$ the structure constants of the gauge algebra. Yang-Mills indices are raised and lowered with the Killing metric, that is assumed to be non-degenerate. We also define
\begin{multieq}[2]
\tor_{MN}{}^{A}&=2\partial_{[M}\tau_{N]}{}^{A}\,,\\
\tore_{MN}{}^{a}&=2\partial_{[M}e_{N]}{}^{a}\,,\\
\tau_{MN}&=\tau_{M}{}^{A}\tau_{N}{}^{B}\eta_{AB}\label{eq:tlong}\,,\\
h_{MN}&=e_{M}{}^{a}e_{N}{}^{b}\delta_{ab}\,,
\end{multieq}
where we define antisymmetrization with weight one, i.e.
\begin{equation}
V_{[A}W_{B]} \equiv \frac{1}{2}(V_AW_B - V_BW_A)\hskip 1truecm  {\rm etc.}
\end{equation}
We adopt the following short-hand notation for contraction of Yang-Mills indices and flat projection of the vector field
\begin{multieq}[3]
v_{\hat{A}\hat{B}}&=v_{\hat{A}I}v_{\hat{B}}^{I}\,,\\
v_{+}^{2}&=v_{+I}v_{+}^{I}\,,\\
v_{-}^{2}&=v_{-I}v_{-}^{I}\,.
\end{multieq}
\\
Longitudinal flat indices could takes also values with respect to lighlike coordinates. These are denoted by $\pm$ and are related to the indices values $A=0,1$ by
\begin{align}
\tau_{M}{}^{\pm}&=\frac{1}{\sqrt{2}}(\tau_{M}{}^{0}\pm \tau_{M}{}^{1})\,,
\end{align}
and analogously for generic longitudinal tensors. Finally we introduce non-relativistic inverse fields $\tau^{M}{}_{A}$ and $e^{M}{}_{a}$ satisfying the usual inverse relations of Newton-Cartan geometry,
\begin{multieqref}[2]{eq:NCinvert}
\tau_{M}{}^{A}e^{M}{}_{a}=\tau^{M}{}_{A}e_{M}{}^{a}&=0\,,\\
e_{M}{}^{a}e^{M}{}_{b}&=\delta^{a}_{b}\,,\\
\tau_{M}{}^{A}\tau^{M B}&=\eta^{A B}\,,\\
\tau_{M}{}^{A}\tau^{N}{}_{A}+e_{M}{}^{a}e^{N}{}_{a}&=\delta_{M}^{N}\,.
\end{multieqref}
Curved indices in $\tore_{MN}{}^{a}$ and $\tor_{MN}{}^{A}$ are converted into flat with inverse vielbein.\\
In the present work we have also used the following redefinition of the vielbein and their inverses
\begin{multieq}[2]
\tnew_{M}{}^{-}&=\tau_{M}{}^{-}\,,\\
\tnew_{M}{}^{+}&=\tau_{M}{}^{+} (1+v_{+-})+e_{M}{}^{a}v_{a-}\,, \phantom{\frac{1}{v_{+-}}}\\
\enew_{M}{}^{a}&=e_{M}{}^{a}\,, \phantom{\frac{1}{v_{+-}}}\\
\tnew^{M}{}_{-}&=\tau^{M}{}_{-}\,,\\
\tnew^{M}{}_{+}&=\frac{1}{1+v_{+-}}\tau^{M}{}_{+}\,,\\
\enew^{M}{}_{a}&=e^{M}{}_{a}-\frac{v_{a-}}{1+v_{+-}}\tau^{M}{}_{+}\,.
\end{multieq}
The hatted Vielbein and their inverses satisfy the same relations as
\autoref{eq:NCinvert}.  In analogy with \autoref{eq:tlong} we denote
\begin{align}
\tnew_{MN}&=\tnew_{M}{}^{A}\tnew_{N}{}^{B}\eta_{AB}\,.
\end{align}

\FloatBarrier
\bibliography{bibliography}{}

\end{document}